\documentclass[10pt,a4paper]{article}

\usepackage[T1]{fontenc}
\usepackage[utf8]{inputenc}
\usepackage{mathptmx}
\usepackage{graphicx}
\usepackage{amsmath,amssymb}
\usepackage{textcomp}
\usepackage{gensymb}
\usepackage[margin=2.2cm]{geometry}
\usepackage{caption}
\usepackage{authblk}
\usepackage[numbers,sort&compress]{natbib}

\usepackage[colorlinks=true,linkcolor=blue,citecolor=blue,urlcolor=blue]{hyperref}

\hypersetup{
  pdftitle={Engineering Polarization Switching in VCSELs with Custom Aperture Shapes},
  pdfauthor={Zifeng Yuan, Dewen Zhang, Hong-Lin Lin, Aaron Danner},
  pdfsubject={Vertical-cavity surface-emitting lasers; polarization switching; photonic Ising machines},
  pdfkeywords={VCSEL, polarization switching, oxide aperture, spin-flip model, optical injection locking, Ising machine, photonic computing}
}

\title{\vspace{-1.2cm}\bfseries\Large Engineering Polarization Switching in VCSELs\\ with Custom Aperture Shapes\vspace{-0.2cm}}

\author[1]{Zifeng Yuan}
\author[1]{Dewen Zhang}
\author[1]{Hong-Lin Lin}
\author[1,*]{Aaron Danner}
\affil[1]{Department of Electrical and Computer Engineering, National University of Singapore, Singapore}
\affil[*]{\href{mailto:adanner@nus.edu.sg}{adanner@nus.edu.sg}}

\date{}

\begin{document}
\maketitle
\thispagestyle{empty}
\pagestyle{empty}

\begin{abstract}
\noindent
We experimentally fabricate and measure vertical-cavity surface-emitting lasers (VCSELs) with custom aperture shapes, demonstrating predictable polarization control and switching, supporting their use in photonic computation schemes like Ising machines with polarization-based information encoding.
\end{abstract}

\section{Introduction}

Vertical-cavity surface-emitting lasers (VCSELs) are widely used semiconductor lasers due to their compact footprints, low power consumption, suitability for on-wafer testing, circular beam profile, and ability to be integrated into large arrays. Optical injection locking, in which a master laser entrains a slave device, is the mechanism underlying much of this utility~\cite{Chang2003}, and has been used to raise modulation bandwidth well beyond the free-running limit~\cite{Zhao2007, Lau2008}. Achieving flexible polarization control and switching on top of this is crucial for many photonics applications, such as Ising computers~\cite{Utsunomiya2011}, where encoding bits or qubits into orthogonal lasing polarization states presents a promising opportunity. Coherent laser networks have realized such machines using multicore fiber lasers~\cite{Babaeian2019}, while VCSEL-based realizations have followed as all-optical annealers with parallel feedback~\cite{Zhang2025} and array-level prototypes~\cite{Lim2024, Zhang2025b}. The same motivation drives integrated optical processors built from other material platforms, such as cascaded interferometer meshes performing matrix--vector products~\cite{Liu2025}. In our previous work, we proposed that such a system would be feasible only if VCSELs could be designed with equal lasing preferences for two orthogonal polarization states~\cite{Gao2024, Loke2023}. However, this remains challenging, as VCSELs typically exhibit a preferred polarization state due to inherent anisotropy~\cite{Gehrsitz2000}.

Polarization behaviour under optical injection has been studied extensively. Orthogonal optical injection can trigger polarization switching accompanied by rich nonlinear dynamics, including bistability and period-doubling routes~\cite{Gatare2006, Perez2011}, and injection-induced switching has been reported for both parallel and orthogonal injection in the 1550~nm window~\cite{Jeong2008, DenisleCoarer2017}. The locking boundaries depend strongly on the injected state of polarization: linear, elliptical and circular injection each produce distinct locking regions and dynamical regimes~\cite{Qader2011, AlSeyab2013, Lin2014}. A recurring conclusion is that the injection power required to flip the polarization state, and the width of the locking range, are set largely by how strongly the device already prefers one axis; reducing that built-in anisotropy through the oxide aperture lowers the required injection power to the microwatt level and widens the locking range~\cite{Yuan2025a}. Polarization stability is likewise sensitive to feedback conditions~\cite{Nazhan2017}, and the orientation of the mesa relative to the crystal axes provides a further degree of freedom~\cite{Yuan2025d}. Beyond polarization, injection locking has been exploited for long-wavelength VCSEL-by-VCSEL links~\cite{Hayat2009}, carrier recovery~\cite{Jignesh2017}, frequency-comb generation~\cite{Prior2016}, spin-polarization modulation~\cite{Yokota2023}, and coupling the emitters of large arrays~\cite{Pfluger2023}.

Controlling the anisotropy at the device level, rather than compensating for it optically, has correspondingly long precedent. Anisotropic transverse cavity geometry was shown early on to fix the polarization axis of a vertical-cavity laser~\cite{Choquette1994b}, and anisotropic cavity geometry combined with injection can induce switching outright~\cite{Tan2012}. Building on this, wafer-scale measurements indicate that tailored apertures give polarization statistics reproducible across large device populations~\cite{Yuan2025c}, which is a prerequisite for the array-level operation these computing schemes require~\cite{Yuan2025e}.

By carefully tailoring the aperture shape of VCSELs, we demonstrate how polarization states and switching can be controlled, achieving stable yet easily switchable polarization states that can be reliably engineered.

\begin{figure}[!ht]
\centering
\includegraphics[width=\textwidth]{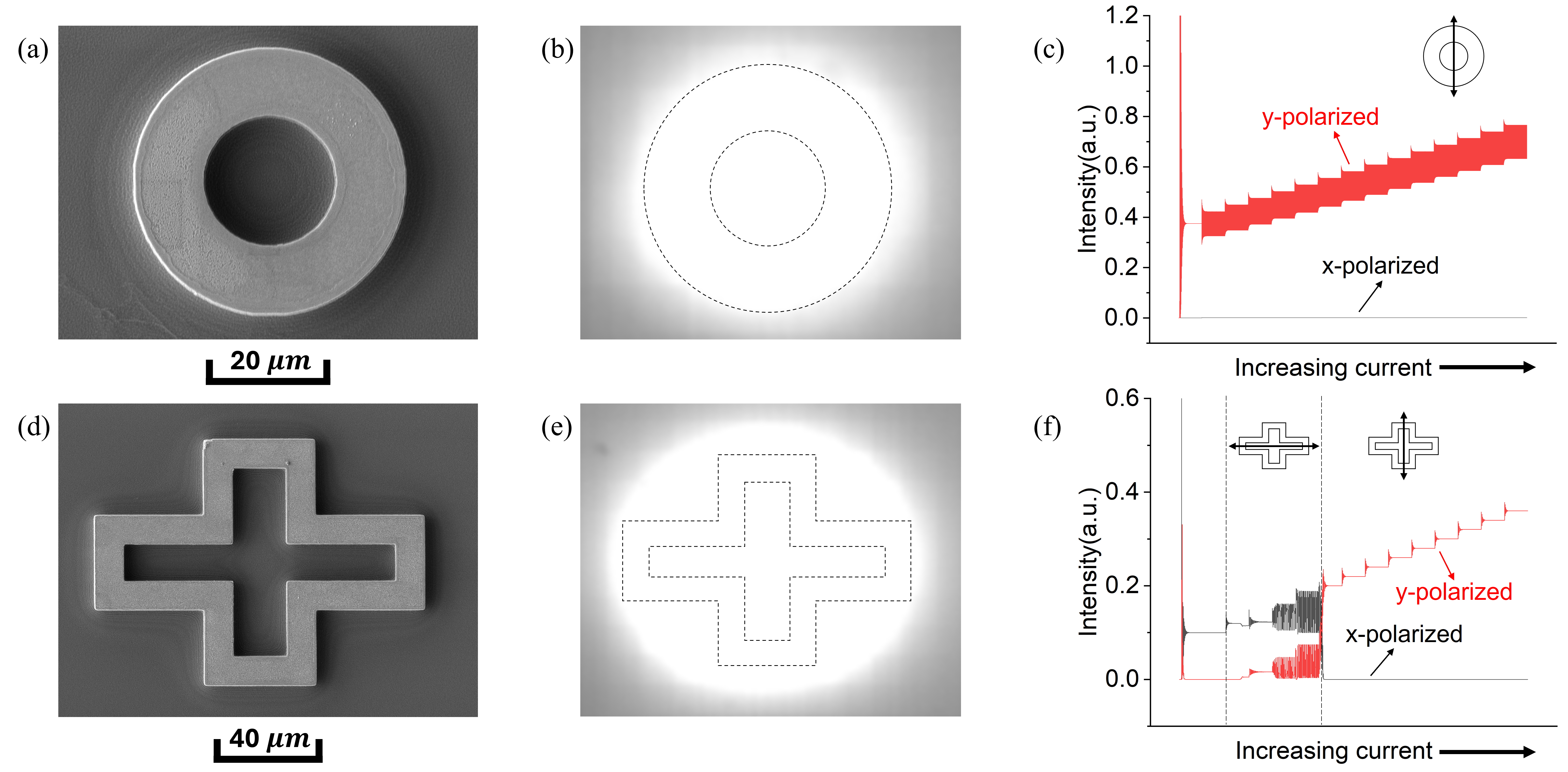}
\caption{(a, d) Scanning electron microscope (SEM) images of a circular VCSEL and a cruciform-shaped VCSEL, respectively. (b, e) Lasing emission images of the circular and cruciform VCSELs. As current increases, numerical results show that (c) the circular VCSEL remains $x$-polarized dominant, while (f) the cruciform VCSEL undergoes a polarization switch from $x$-polarized to $y$-polarized. Fig.~2 shows experimental confirmation of this behavior.}
\label{fig:fig1}
\end{figure}

\section{Results}

VCSELs with various custom aperture shapes were fabricated and tested. Fig.~\ref{fig:fig1} shows scanning electron microscopy (SEM) images (a, d) and lasing emission images (b, e) of fabricated circular and cruciform VCSELs, respectively. VCSELs typically exhibit two dominant linear polarization modes aligned with crystal axes, but inherent gain anisotropy is expected due to factors such as crystal strain~\cite{Gehrsitz2000}. The cruciform design is intended to modify the gain in two orthogonal directions~\cite{Choquette1994b}. The Spin Flip Model (SFM) provides a theoretical framework for analyzing the polarization bistability of $x$- and $y$-polarized VCSEL modes~\cite{Martin1997, AlSeyab2011}. By incorporating the gain anisotropy of circular and cruciform VCSELs into the SFM, we observe distinct polarization switching behavior, as illustrated in Fig.~\ref{fig:fig1}(c) and Fig.~\ref{fig:fig1}(f). For the circular VCSEL, the polarization remains $x$-polarized without switching as the current increases, a behavior attributed to natural anisotropy inherent in VCSELs, as shown in Fig.~\ref{fig:fig1}(c). In contrast, the cruciform VCSEL exhibits a polarization switch from $x$-polarized to $y$-polarized as current increases, as illustrated in Fig.~\ref{fig:fig1}(f).

\begin{figure}[!ht]
\centering
\includegraphics[width=\textwidth]{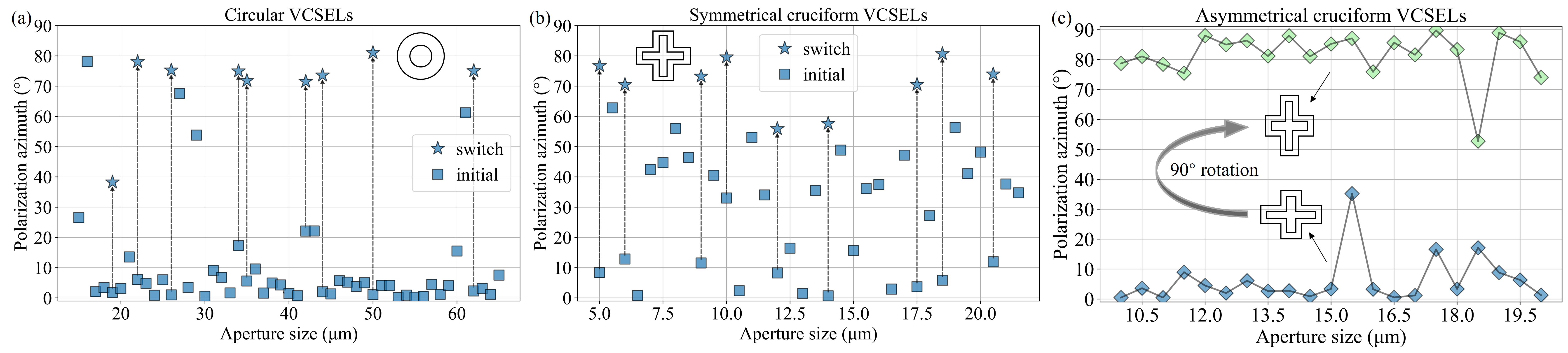}
\caption{Experimental polarization azimuth results for various VCSEL aperture designs, with aperture size on the $x$-axis and average azimuth on the $y$-axis. (a) Circular VCSELs, which predominantly lase in the $x$-polarized state and seldom switch polarization. (b) Symmetrical cruciform VCSELs, showing an increasing average polarization azimuth and more frequent polarization switching. (c) Initial polarization states of unrotated and 90\degree-rotated asymmetrical cruciform VCSELs, where the arm length ratio is 0.7.}
\label{fig:fig2}
\end{figure}

Experimental results are shown in Fig.~\ref{fig:fig2}(a) and Fig.~\ref{fig:fig2}(b) for circular and symmetrical cruciform VCSELs, respectively. The initial polarization state (at low current, just above threshold) is indicated by a square, while the switched polarization state (after increasing current) is marked by a star, if observed. Most circular VCSELs lase $x$-polarized, while symmetrical cruciform VCSELs exhibit higher average polarization azimuths and a greater proportion of devices with polarization switching. (The rotation angles are defined relative to the wafer axis, with the long arm initially aligned along the $x$-axis.) Next, we investigate VCSELs with asymmetrical cruciform apertures, where the length ratio of the two arms is set to 0.7, intuitively chosen based on prior studies~\cite{Tan2012}. Fig.~\ref{fig:fig2}(c) illustrates initial polarization states of unrotated and 90\degree-rotated VCSELs. Most unrotated VCSELs lase $x$-polarized, while 90\degree-rotated VCSELs lase $y$-polarized. This demonstrates that the polarization state of VCSELs can be manipulated with the asymmetrical cruciform design. Rotating the orientation angle of the cruciform leads to predictable shifts in polarization states, presenting an opportunity for precise engineering of polarization behavior.

\section{Conclusion}

In conclusion, by fabricating and analyzing VCSELs with custom aperture shapes, we demonstrate that polarization control and switching can be effectively engineered. Circular VCSELs exhibit stable polarization states, while cruciform designs allow for manipulation of polarization behavior, including switching and orientation-dependent states. These findings highlight the potential of VCSELs with engineered apertures for photonic computation applications, such as dedicated Ising machine architectures~\cite{Zhang2025Patent}, where polarization-based information encoding is crucial, as well as for co-integrated photonic components~\cite{Lim2025}. Reliable polarization encoding may further support emerging benchmarks and applications in computational imaging~\cite{lin2025rgb, Teng2025} and in machine learning more broadly~\cite{Chen2026hypo, chen2025auto, Du2026}.

\subsection*{Acknowledgments}

This work was supported by the National Research Foundation, Singapore, under its Competitive Research Programme (NRF CRP24-2020-0003) and by both the National Research Foundation, Singapore, and A*STAR under the Quantum Engineering Programme (NRF 2021-QEP2-02-P12).

\renewcommand{\refname}{References}
\small
% Numbered in order of first appearance in the text.
\bibliographystyle{unsrt}
\bibliography{references}

\begin{thebibliography}{10}

\bibitem{Chang2003}
C.~H. Chang, L.~Chrostowski, and C.~J. Chang-Hasnain.
\newblock Injection locking of {VCSELs}.
\newblock {\em IEEE Journal of Selected Topics in Quantum Electronics},
  9(5):1386--1393, 2003.

\bibitem{Zhao2007}
X.~Zhao, D.~Parekh, E.~K. Lau, et~al.
\newblock Novel cascaded injection-locked 1.55-$\mu$m {VCSELs} with 66 {GHz}
  modulation bandwidth.
\newblock {\em Optics Express}, 15(22):14810--14816, 2007.

\bibitem{Lau2008}
E.~K. Lau, X.~Zhao, H.~K. Sung, et~al.
\newblock Strong optical injection-locked semiconductor lasers demonstrating
  $>$100-{GHz} resonance frequencies and 80-{GHz} intrinsic bandwidths.
\newblock {\em Optics Express}, 16(9):6609--6618, 2008.

\bibitem{Utsunomiya2011}
S.~Utsunomiya, K.~Takata, and Y.~Yamamoto.
\newblock Mapping of {Ising} models onto injection-locked laser systems.
\newblock {\em Optics Express}, 19(19):18091--18108, 2011.
\newblock doi:10.1364/OE.19.018091.

\bibitem{Babaeian2019}
M.~Babaeian, D.~T. Nguyen, V.~Demir, et~al.
\newblock A single shot coherent {Ising} machine based on a network of
  injection-locked multicore fiber lasers.
\newblock {\em Nature Communications}, 10(1):3516, 2019.
\newblock doi:10.1038/s41467-019-11548-4.

\bibitem{Zhang2025}
D.~Zhang, Z.~Yuan, T.~X. Hoang, W.~Fu, C.~E. Png, S.~T. Lim, and A.~Danner.
\newblock All-optical scalable and programmable {VCSEL}-based {Ising} annealer
  with parallel feedback.
\newblock {\em Optics Express}, 33(11):22119--22131, 2025.
\newblock doi:10.1364/OE.546402.

\bibitem{Lim2024}
S.~T. Lim, D.~Zhang, Z.~Yuan, T.~X. Hoang, C.~E. Png, and A.~Danner.
\newblock Quantum-inspired {VCSEL}-based optical parallel feedback system for
  large-scale combinatorial optimization.
\newblock In {\em 2024 IEEE Opto-Electronics and Communications Conference
  (OECC)}, pages 1--3, Melbourne, Australia, 2024.
\newblock doi:10.1109/OECC54135.2024.10975547.

\bibitem{Zhang2025b}
D.~Zhang, Z.~Yuan, T.~X. Hoang, C.~E. Png, S.~T. Lim, and A.~Danner.
\newblock A {VCSEL}-based optical {Ising} computer with parallel feedback
  system for large-scale optimization.
\newblock In {\em CLEO: Applications and Technology}, Technical Digest Series,
  page JPS200\_182. Optica Publishing Group, 2025.

\bibitem{Liu2025}
Y.~Liu, P.~Aashna, Z.~Yuan, W.~Fu, M.~Yang, Y.~Yan, L.~Qi, M.~Zhao, and
  A.~Danner.
\newblock A lithium niobate cascaded {MMI}-based arbitrary matrix calculator
  for convolutional neural networks.
\newblock {\em Optics \& Laser Technology}, 192:113740, 2025.
\newblock doi:10.1016/j.optlastec.2025.113740.

\bibitem{Gao2024}
Y.~Gao, G.~Chen, L.~Qi, W.~Fu, Z.~Yuan, and A.~J. Danner.
\newblock Photonic {Ising} machines for combinatorial optimization problems.
\newblock {\em Applied Physics Reviews}, 11(4):041307, 2024.
\newblock doi:10.1063/5.0216656.

\bibitem{Loke2023}
B.~Loke, Z.~Yuan, S.~T. Lim, and A.~Danner.
\newblock Linear polarization state encoding for {Ising} computing with
  optically injection-locked {VCSEL}s.
\newblock {\em Journal of Optical Microsystems}, 4(1):014501, 2024.
\newblock doi:10.1117/1.JOM.4.1.014501.

\bibitem{Gehrsitz2000}
S.~Gehrsitz, F.~K. Reinhart, C.~Gourgon, et~al.
\newblock The refractive index of {Al}$_x${Ga}$_{1-x}${As} below the band gap:
  Accurate determination and empirical modeling.
\newblock {\em Journal of Applied Physics}, 87(11):7825--7837, 2000.
\newblock doi:10.1063/1.373462.

\bibitem{Gatare2006}
I.~Gatare, M.~Sciamanna, J.~Buesa, et~al.
\newblock Nonlinear dynamics accompanying polarization switching in
  vertical-cavity surface-emitting lasers with orthogonal optical injection.
\newblock {\em Applied Physics Letters}, 88(10):101106, 2006.

\bibitem{Perez2011}
P.~P{\'e}rez, A.~Quirce, L.~Pesquera, et~al.
\newblock Polarization-resolved nonlinear dynamics induced by orthogonal
  optical injection in long-wavelength {VCSELs}.
\newblock {\em IEEE Journal of Selected Topics in Quantum Electronics},
  17(5):1228--1235, 2011.

\bibitem{Jeong2008}
K.~H. Jeong, K.~H. Kim, S.~H. Lee, et~al.
\newblock Optical injection-induced polarization switching dynamics in
  1.5-$\mu$m wavelength single-mode vertical-cavity surface-emitting lasers.
\newblock {\em IEEE Photonics Technology Letters}, 20(10):779--781, 2008.

\bibitem{DenisleCoarer2017}
F.~Denis le~Coarer, A.~Quirce, P.~P{\'e}rez, et~al.
\newblock Injection locking and polarization switching bistability in a 1550 nm
  {VCSEL} subject to parallel optical injection.
\newblock {\em IEEE Journal of Selected Topics in Quantum Electronics},
  23(6):1801510, 2017.

\bibitem{Qader2011}
A.~A. Qader, Y.~Hong, and K.~A. Shore.
\newblock Lasing characteristics of {VCSELs} subject to circularly polarized
  optical injection.
\newblock {\em Journal of Lightwave Technology}, 29(24):3804--3809, 2011.

\bibitem{AlSeyab2013}
R.~Al-Seyab, K.~Schires, A.~Hurtado, et~al.
\newblock Dynamics of {VCSELs} subject to optical injection of arbitrary
  polarization.
\newblock {\em IEEE Journal of Selected Topics in Quantum Electronics},
  19(4):1700512, 2013.

\bibitem{Lin2014}
H.~Lin, P.~P{\'e}rez, A.~Valle, et~al.
\newblock Investigation of elliptically polarized injection locked states in
  {VCSELs} subject to orthogonal optical injection.
\newblock {\em Optics Express}, 22(5):4880--4885, 2014.

\bibitem{Yuan2025a}
Z.~Yuan, D.~Zhang, L.~Shi, Y.~Liu, and A.~Danner.
\newblock Enhanced polarization locking in {VCSELs}.
\newblock {\em Applied Physics Letters}, 126(15):151101, 2025.
\newblock doi:10.1063/5.0259836.

\bibitem{Nazhan2017}
S.~Nazhan and Z.~Ghassemlooy.
\newblock Polarization switching dependence of {VCSEL} on variable polarization
  optical feedback.
\newblock {\em IEEE Journal of Quantum Electronics}, 53(4):2400807, 2017.

\bibitem{Yuan2025d}
Z.~Yuan, W.~Shan, T.~Chen, B.~Lin, and A.~Danner.
\newblock Mesa orientation engineering for polarization locking in {VCSELs}.
\newblock In {\em 2025 IEEE Photonics Conference (IPC)}, pages 1--2, Singapore,
  Singapore, 2025.
\newblock doi:10.1109/IPC65510.2025.11282171.

\bibitem{Hayat2009}
A.~Hayat, A.~Bacou, A.~Rissons, et~al.
\newblock Long wavelength {VCSEL}-by-{VCSEL} optical injection locking.
\newblock {\em IEEE Transactions on Microwave Theory and Techniques},
  57(7):1850--1858, 2009.

\bibitem{Jignesh2017}
J.~Jignesh, B.~Corcoran, J.~Schr{\"o}der, et~al.
\newblock Polarization independent optical injection locking for carrier
  recovery in optical communication systems.
\newblock {\em Optics Express}, 25(18):21216--21228, 2017.

\bibitem{Prior2016}
E.~Prior, C.~De Dios, R.~Criado, et~al.
\newblock Dynamics of dual-polarization {VCSEL}-based optical frequency combs
  under optical injection locking.
\newblock {\em Optics Letters}, 41(17):4083--4086, 2016.

\bibitem{Yokota2023}
N.~Yokota, K.~Ikeda, and H.~Yasaka.
\newblock Observation of spin polarization modulation responses of
  injection-locked vertical-cavity surface-emitting lasers.
\newblock {\em IEICE Electronics Express}, 20(8):20230057, 2023.

\bibitem{Pfluger2023}
M.~Pfl{\"u}ger, D.~Brunner, T.~Heuser, et~al.
\newblock Injection locking and coupling the emitters of large {VCSEL} arrays
  via diffraction in an external cavity.
\newblock {\em Optics Express}, 31(5):8704--8713, 2023.

\bibitem{Choquette1994b}
K.~D. Choquette and R.~E. Leibenguth.
\newblock Control of vertical-cavity laser polarization with anisotropic
  transverse cavity geometries.
\newblock {\em IEEE Photonics Technology Letters}, 6(1):40--42, 1994.
\newblock doi:10.1109/68.265883.

\bibitem{Tan2012}
M.~P. Tan, A.~M. Kasten, T.~A. Strand, et~al.
\newblock Polarization switching in vertical-cavity surface-emitting lasers
  with anisotropic cavity geometry and injection.
\newblock {\em IEEE Photonics Technology Letters}, 24(9):745--747, 2012.
\newblock doi:10.1109/LPT.2012.2187330.

\bibitem{Yuan2025c}
Z.~Yuan, D.~Zhang, Y.~Gao, L.~Qi, W.~Fu, and A.~Danner.
\newblock Large-scale fabrication and analysis of polarization behavior in
  {VCSELs} with tailored apertures.
\newblock {\em Journal of Lightwave Technology}, 43(14):6819--6827, 2025.
\newblock doi:10.1109/JLT.2025.3568130.

\bibitem{Yuan2025e}
Z.~Yuan, T.~Chen, D.~Zhang, Y.~Gao, W.~Shan, B.~Lin, and A.~Danner.
\newblock Tailored polarization-switchable {VCSEL} arrays for photonic {Ising}
  computing.
\newblock {\em Applied Physics Letters}, 127(22):221102, 2025.
\newblock doi:10.1063/5.0291349.

\bibitem{Martin1997}
J.~Martin-Regalado, F.~Prati, M.~San Miguel, and N.~B. Abraham.
\newblock Polarization properties of vertical-cavity surface-emitting lasers.
\newblock {\em IEEE Journal of Quantum Electronics}, 33(5):765--783, 1997.
\newblock doi:10.1109/3.572151.

\bibitem{AlSeyab2011}
R.~Al-Seyab, K.~Schires, N.~A. Khan, et~al.
\newblock Dynamics of polarized optical injection in 1550-nm {VCSELs}: Theory
  and experiments.
\newblock {\em IEEE Journal of Selected Topics in Quantum Electronics},
  17(5):1242--1249, 2011.
\newblock doi:10.1109/JSTQE.2011.2138683.

\bibitem{Zhang2025Patent}
D.~Zhang, Z.~Yuan, A.~J. Danner, S.~T. Lim, and T.~X. Hoang.
\newblock {VCSEL}-based all-optical {Ising} machine, 2025.
\newblock US Patent Application 19/068,345.

\bibitem{Lim2025}
S.~T. Lim, T.~Y.~L. Ang, Z.~Yuan, T.~X. Hoang, D.~Zhang, C.~E. Png, A.~J.
  Danner, and G.~Alagappan.
\newblock Advancing detector shielding with thermo-optic defocusing in {PMMA}
  integrated on silicon nitride.
\newblock In {\em Proceedings of the International Conference on Photonics,
  Optics and Laser Technology (PHOTOPTICS)}, pages 92--94, 2025.

\bibitem{lin2025rgb}
Beibei Lin, Zifeng Yuan, and Tingting Chen.
\newblock {RGB}-to-polarization estimation: A new task and benchmark study.
\newblock In {\em Advances in Neural Information Processing Systems (NeurIPS)},
  2025.
\newblock arXiv:2505.13050.

\bibitem{Teng2025}
Z.~Teng, T.~Chen, B.~Lin, Z.~Yuan, X.~Li, X.~Zhang, and S.~Zhang.
\newblock {RaindropGS}: A benchmark for {3D} {Gaussian} splatting under
  raindrop conditions.
\newblock {\em arXiv preprint arXiv:2510.17719}, 2025.

\bibitem{Chen2026hypo}
T.~Chen, B.~Lin, Z.~Yuan, Q.~Zou, H.~He, Y.-S. Ong, A.~Goyal, and D.~Liu.
\newblock {HypoSpace}: Evaluating {LLM} creativity as set-valued hypothesis
  generators under underdetermination.
\newblock In {\em Proceedings of the 43rd International Conference on Machine
  Learning (ICML)}, 2026.
\newblock arXiv:2510.15614.

\bibitem{chen2025auto}
Tingting Chen, Beibei Lin, Srinivas Anumasa, Vedant Shah, Zifeng Yuan, Qiran
  Zou, Anirudh Goyal, and Dianbo Liu.
\newblock {Auto-Discovery-Bench}: Diagnosing structured state tracking in
  oracle-guided discovery.
\newblock {\em arXiv preprint arXiv:2502.15224}, 2025.

\bibitem{Du2026}
W.~Du, Z.~Yuan, T.~Chen, F.~Ke, B.~Lin, and S.~Zhang.
\newblock {WeatherReasonSeg}: A benchmark for weather-aware reasoning
  segmentation in visual language models.
\newblock In {\em European Conference on Computer Vision (ECCV)}, 2026.
\newblock arXiv:2603.17680.

\end{thebibliography}

\end{document}